\relax
\documentclass[letterpaper]{article} % DO NOT CHANGE THIS
\usepackage{aaai21}  % DO NOT CHANGE THIS
\usepackage{times}  % DO NOT CHANGE THIS
\usepackage{helvet} % DO NOT CHANGE THIS
\usepackage{courier}  % DO NOT CHANGE THIS
\usepackage[hyphens]{url}  % DO NOT CHANGE THIS
\usepackage{graphicx} % DO NOT CHANGE THIS
\usepackage{natbib}  % DO NOT CHANGE THIS AND DO NOT ADD ANY OPTIONS TO IT
\usepackage{caption} % DO NOT CHANGE THIS AND DO NOT ADD ANY OPTIONS TO IT
\usepackage[ruled,vlined]{algorithm2e}
\usepackage{amsmath}
\usepackage{amssymb}
\usepackage{amsthm}
\usepackage{xspace}
\newcommand{\graph}[1] {\ensuremath{\mathcal {#1}}}
\newcommand{\vars}[1] {\ensuremath{\mathbf {#1}}}
\newcommand{\var}[1] {\ensuremath{#1}}

\newcommand{\hz} {{\ensuremath{\mathcal {H}_\mathbf{z}}\!}\;}

\newcommand\mydots{\ifmmode\ldots\else\makebox[1em][c]{.\hfil.\hfil.}\fi}

\newtheorem{definition}{Definition}

\makeatletter
\newcommand{\shorteq}{%
  \settowidth{\@tempdima}{-}% Width of hyphen
  \resizebox{\@tempdima}{\height}{=}%
}
\def\Equal{\texttt{=}}

\title{Learning Adjustment Sets from Observational and Limited Experimental Data}
\author{

    Sofia Triantafyllou \\
    Gregory Cooper \\
}
\affiliations{
    Department of Biomedical Informatics, University of Pittsburgh
}
\begin{document}

\maketitle

\begin{abstract}
We can estimate causal effects from observational data if an appropriate set of covariates (an adjustment set) can be identified, which removes confounding bias; however, such a set is often not identifiable from observational data alone. Experimental data allow unbiased causal effect estimation, but are typically limited in sample size and can therefore yield estimates of high variance. Moreover, experiments are often performed on a different (specialized) population than the population of interest. In this work, we introduce a method that combines large observational and limited experimental data to identify adjustment sets and improve the estimation of causal effects for a target population. The method scores an adjustment set by calculating the marginal likelihood for the experimental data given an observationally-derived causal effect estimate, using a putative adjustment set. The method can make inferences that are not possible using constraint-based methods. We show that the method can improve causal effect estimation, and  can  make additional inferences when compared to state-of-the-art methods. 
\end{abstract}
Covariate adjustment is the main method for estimating causal effects from observational data. There is a lot of work on  identifying the correct sets for covariate adjustment in the fields of potential outcomes and causal graphs. For the latter, sound and complete graphical criteria have recently been proven \citep{van2014constructing, shpitser2012validity}. These criteria allow the identification of all the variable sets that lead to unbiased estimates of post-interventional probabilities through covariate adjustment \emph{when the causal graph is known}. Unfortunately, the true causal graph is  often unknown. Causal discovery methods try to identify the causal graph for a set of variables based on the causal Markov and  faithfulness assumptions \citep{spirtes2000causation}. Often, multiple graphs fit the data equally well and are called Markov equivalent (ME). Thus, the correct sets for covariate adjustment are often  not uniquely identifiable from observational data alone. In contrast, experimental data are the gold standard for estimating unbiased causal effects, but are often limited in terms of sample size, leading to estimates with high variance. Moreover,  experiments are often performed on a specialized population (e.g., a particular age distribution) and the effect estimated from the experimental data does not apply directly to the observational population.

\begin{figure}[t]\label{example1}
\centering
\includegraphics[width =\columnwidth]{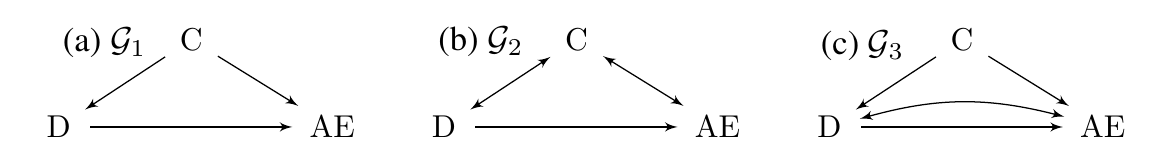}
\caption{\label{fig:example1}\textbf{Markov equivalent graphs imply different IDs} In $\graph G_1$ \var C is an adjustment set for $D, AE$ and $P_{D}(AE) =\sum_c P(AE|D, c)P(c)$. In $\graph G_2$, $\emptyset$ is an adjustment set for $X, Y$ and $P_{D}(AE) \Equal P(AE|D)$. In $\graph G_3$, $P_{D}(AE)$  is not identifiable from observational data. The  graphs are indistinguishable based on the conditional (in) dependencies in observational data over $\{\var D, \var AE, \var C\}$ and experimental data over  $\{\var D, \var AE\}$, but entail different expressions for $P_{D}(AE)$.} 
\end{figure}
We introduce a method for combining observational and limited experimental data (that can be extracted from a publication) to find an adjustment set, if one exists, or identify that none exists. The method is motivated by the following common scenario: Assume that a researcher is interested in quantifying an adverse effect ($AE$) of a drug ($D$) on a population and has access to a large collection of electronic health records (EHR) of patients who take the drug or not, along with some covariates. The researcher also has the published results of a randomized control trial (RCT)  that reports the estimated causal effect $\hat P(AE|do(D))$ (or $\hat P_D(AE)$) and deems it significant; thus, $D$ \emph{causes} $AE$. The researcher suspects that this causal relationship is  confounded by another condition ($C$), not included in the RCT, that is highly correlated with both $D$ and $AE$ in the observational data. She wants to know if $C$ is an adjustment set for $AE$ and $D$ in the observational data. While the RCT already provides an estimate for $P_D(AE)$, using the more plentiful observational data with covariate adjustment can improve this estimate. Moreover, if $C$ is an adjustment set, it can be used to provide a more personalized prediction $P_D(AE|C)$. This cannot be estimated from the RCT alone, because the RCT does not include $C$. 

Alternatively, the researcher may also believe that the distribution of $C$ in the RCT is different than the one in their observational population, based on some reported marginals in the RCT paper (RCT publications typically report marginal distributions of some background covariates). In this case, the RCT estimate  $\hat P_D(AE)$ may not be accurate for the observational population. The researcher wants to know if they can estimate it by adjusting for $C$ in the observational data.

Fig. \ref{fig:example1} shows possible graphs for the first scenario. The graphs cannot be distinguished based on (in) dependence constraints in the EHR and RCT data (\var C is not measured in the RCT). However, they imply different ways for computing the Interventional Distribution (ID) $P_D(AE)$  from observational data. For two of these graphs ($\graph G_1$ and $\graph G_2$), the ID can be computed from the observational data based on the appropriate covariate adjustment, which differs for the two graphs. 

In this work, we present a Bayesian method that combines observational data and limited experimental data, to identify if an adjustment set exists. The method looks in the observational covariates for a  set that leads to an unbiased ID estimate. If such a set exists, the method uses it to reduce the variance of the ID estimate (or, if the observational and experimental data come from different populations, to get an unbiased estimate for the ID in the observational population.)

The proposed method scores an adjustment set by calculating the marginal likelihood for the experimental data given an observationally-derived causal effect estimate, using the putative adjustment set. The method can identify valid adjustment sets, even when these are not uniquely identified from the ME class of graphs consistent with the observation and experimental data: For example, the method can identify that $C$ is an adjustment set for $D, AE$ in Fig. \ref{fig:example1}(a), even though $C$ is not measured in the experimental data. In addition, the method can identify that there is no adjustment set in the observational data. To our knowledge, this method is the first one described in the literature that can make this inference.  

\section{Preliminaries}
We use the framework of Semi-Markov Causal Models (SMCMs). We assume the reader is familiar with causal graphical models and related terminology.  We use the terms node and variable interchangeably. We use bold to denote variable sets, uppercase letters to denote single variables, and lowercase letters to denote variable values.  If we know the causal SMCM \graph G, a hard intervention of where a treatment $X$ is set to $x$  can be represented with the do-operator, $do(X\Equal x)$. The ID of an outcome $Y$ given $do(X\Equal x)$ is denoted $P(Y|do(X\Equal x))$ or $P_x(Y)$. In the corresponding SMCM, this is equivalent to removing all incoming edges into $\var X$, while keeping all other mechanisms intact. % We use  $\graph G_{\overline{X}}$ to denote the graph stemming from \graph G after removing edges into $X$. %We use $\graph G_{\underline{X}}$ to denote the graph stemming from \graph G after removing edges out of \var X.

One way to estimate  $P_X(Y)$ from observational data is with \textbf{covariate adjustment}. The goal of this process is to control for confounding bias, without introducing additional bias (e.g., m-bias \citep{greenland2003quantifying}). Thus, adjustment amounts to selecting a proper set of variables \vars Z and ``adjusting" for their effect to obtain the ID:
\begin{equation}\label{eq:do-calc}
P_Y(x) =P(Y|do(X\Equal x))= \sum_\vars z P(Y|x, \vars z)P(\vars z)\quad \forall x, y \end{equation} Eq. \ref{eq:do-calc} is called the \textbf{adjustment formula}, and set \vars Z is an \textbf{adjustment set} for $X$ and $Y$.
If we know the causal SMCM \graph G, we can identify all valid adjustment sets using a sound and complete graphical criterion, called the \textbf{adjustment criterion}  \citep{shpitser2012validity}.
% \begin{definition}\label{def:adjustment_criterion}
% $\mathbf Z$ satisfies the adjustment criterion relative to $(X,Y)$ in \graph G if (a) no node in \vars Z is a descendant in \graph G of any node that lies on a causal\footnote{In \citep{shpitser2012validity}, the authors use the term a proper causal path to allow for sets of exposures and outcomes. For singletons \var X and \var Y this is identical to a causal path $\var X \dashrightarrow \var Y$.} path from \var X to \var Y and (b) all non-causal paths in \graph G from \var X to \var Y are blocked by \vars Z.
% \end{definition}
%The criterion is sound, meaning that any set \vars Z that satisfies the adjustment criterion for $(X, Y)$ is an adjustment set for  $(X, Y)$ in all distributions that induce \graph G. It is also complete, meaning that, for any \vars Z that do not satisfy the adjustment criterion, there exists a  distribution \graph P inducing \graph G where \vars Z is not an adjustment set for $(X, Y)$. 

\section{Scoring Adjustment Sets}
The adjustment criterion allows us to identify all  adjustment sets for \var X and \var Y (if any) in an SMCM \graph G. We can then use an adjustment set to estimate the ID $P_X(\var Y)$ from the pre-intervention distribution $P(\vars V)$. Since we often do not know the graph \graph G,  we are interested in reverse engineering the adjustment sets for (\var X, \var Y) using the empirical observational JPD $\hat P(\vars V)$, when an empirical $\hat P_X(\var Y)$ is also available.

\begin{algorithm}[t]
\LinesNumbered
\SetKwInOut{Input}{input}
\SetKwInOut{Output}{output}
\SetKwFunction{FindPAG}{FindPAG}
\SetKwFunction{scoreExp}{scoreExp}
\SetKwFunction{EstProbObs}{EstProbObs}
\SetKwFunction{LearnBN}{LearnBN}
\Input{$Y,D_{obs},D_{exp}=\{D_{x}\}_{x\in X},niters$}
\Output{Adjustment set $\vars Z^\star$,  estimate $\bar P_{X}(Y)$}
$\graph B_{post}=\langle\graph B, f(\vars\theta_{i|pa_i}|\graph B, D_{obs})\rangle\leftarrow \LearnBN(D_{obs})$\;
$\vars Z_{XY} \leftarrow$ Variables associated with $X$ and $Y$\;
\ForEach{subset \vars  Z of $\vars Z_{XY}$ and $\nexists$}{
$P(\hz|D_{obs}) \leftarrow \EstProbObs(\vars Z, D_{obs})$\;
$\{P(D_{exp}|D_{obs}, \hz),  p\vars Z\}\leftarrow\prod_{x}\scoreExp(x, Y,  \vars Z,D_{obs}, D_{x}, \graph B_{post}, niters)$}
$\vars Z^\star\leftarrow argmax_\vars z P(D_{exp}|D_{obs}, \hz)P(\hz|D_{obs})$\;
$\bar P_{X}(Y)\leftarrow p\vars Z^\star$\;
\caption{findAdjustmentSet (FAS) \label{algo:findBestAdj}}
\end{algorithm}
We assume the  following setting: There exists a SMCM \graph G over a set of variables \vars V and a JPD \graph P over the same variables such that \graph G and \graph P are faithful to each other. The variables include a treatment \var X and an outcome \var Y caused by \var X. We present our results for discrete variables and a multinomial distribution, but the results can be extended  to other distributions for which marginal likelihoods can be computed in closed form or approximated. We assume we have: 
\begin{itemize}
\item Observational data $D_{obs}$ measuring \vars V, over $N$ samples.
\item Experimental data $D_{exp} =\{D_x\}_x\in X$. Each $D_x$ consists of an estimate of $\hat P_{X=x}(Y)$, and the corresponding sample size $N_{x}$.
\end{itemize}
It is common in biology and medicine that the information in $D_{exp}$ is  included in the publication that presents an RCT. Typically, the publication also includes estimates for the marginal distributions for a set of additional covariates $\vars V_{exp}$. These distributions can be used to handle cases where $D_{exp}$ is collected in a population different than $D_{obs}$. We discuss this in the section ``Dealing with selection in the experimental data." We first present the method for cases where $D_{obs}$ and $D_{exp}$ come from the same population. 
%Notice that $D_{exp}$ does not have information about the joint distribution of $P(X, Y, \vars V_{exp})$, and thus, cannot be queried for conditional independencies.

\begin{algorithm}[t]
\LinesNumbered
\SetKwInOut{Input}{input}
\SetKwInOut{Output}{output}
\SetKwFunction{SampleBN}{SampleBN}
\SetKwFunction{BayesInf}{BayesInf}
\SetKwFunction{ScoreDe}{$ScoreD_e$}
\Input{$X, Y,\vars Z,D_{obs},D_{x},\graph B_{post}, niters$}
\Output{$P(D_{x}| D_{obs}, \hz)$, $\bar P_{x}(Y)$}
\If{$\vars Z\neq\nexists$}
{\ForEach{$iter =1, \dots,niters $}{
Sample $\widetilde\theta_{i|pa_i}\sim f(\theta_{i|pa_i}|\graph B, D_{obs}) $\;
${\widetilde\theta_{Y|x, \vars z}, \widetilde\theta_\vars z}\leftarrow \BayesInf(\graph B, \widetilde\theta_{i|pa_i})$\;
$\widetilde \theta_{Y_x}(iter)\leftarrow \sum_\vars z \widetilde\theta_{Y|x, \vars z}\widetilde\theta_\vars z$\;
$\widetilde pZ(iter) \leftarrow \prod_y\widetilde \theta_{y_x}(iter)^{N^y_{x}}$\;
    }
    $P(D_{x}|D_{obs}, \hz)\leftarrow  \overline{\widetilde pZ}$, $\bar P_{x}(Y) \leftarrow \overline{\widetilde \theta_{Y_x}}$\; }
\Else{
$P(D_{x}|D_{obs},\mathcal H_{\not\exists}) \leftarrow
\prod_y{\frac{\prod_i\Gamma (N^y_{x})}{\Gamma(N_{x}+|Y|)}}$\;$\bar P_x(Y) \leftarrow  \hat P_{x}(Y)$\;}
\caption{ScoreExp}\label{algo:pde}
\end{algorithm}
 We introduce a Bayesian method, presented in Algorithms \ref{algo:findBestAdj} and \ref{algo:pde}. Intuitively, our method is based on the following observation:  Different causal graphs, consistent with the conditional (in)dependence constraints in the data, may entail different adjustment sets for $(X, Y)$, which in turn may lead to different predicted IDs $\bar P_X(Y)$. In addition, there may be cases where no adjustment set exists among the set of observed variables, and therefore the observational data cannot be used to identify the  ID through covariate adjustment. By (implicitly) comparing  $\bar P_X(Y)$ and the estimate $\hat P_X(Y)$ in the experimental data, we can identify sets that are more probable to be adjustment sets for $(X, Y)$, and use them to improve the estimate for $P_X(Y)$. 
% To do so, we need something similar to faithfulness for the adjustment criterion. Specifically, we will assume that the adjustment sets for (\var X, \var Y) are exactly those for which the adjustment criterion holds. We call this assumption \textbf{adjustment faithfulness}:
% \begin{definition}Let \graph G be a causal SMCM and \graph P a distribution faithful to \graph G. Then for all disjoint sets of variables \var X, \var Y, \vars Z, \vars Z is an adjustment set for (\var X, \var Y) in \graph P  (according to Equation 1) only if \vars Z satisfies the adjustment criterion for (\var X, \var Y) in  \graph G.
% \end{definition}
% This assumption rules out distributions where \vars Z is an adjustment set for (\var X, \var Y) without satisfying the adjustment criterion. In the rest of this paper, 
We use a binary variable \hz to denote that $\vars Z$ is an adjustment set for $(X, Y)$ (thus, \hz is true if \vars Z is an adjustment set for $X, Y$). It is also possible that no adjustment set exists among \vars V. We denote this hypothesis as $\mathcal H_{\not \exists}$. Note that this is different than  $\mathcal H_{\emptyset}$, which states that the empty set is an adjustment set. $\mathcal H_{\not \exists}$ complements the space of possible hypotheses $\hz$ with respect to the adjustment criterion. 
%To denote that $\vars Z$ is an adjustment set, we use the following notation: Let \hz be a binary variable that is true if $\vars Z$ is an adjustment set for $(X, Y)$, and let $\graph G\vdash \hz$ denote that $\vars Z$ satisfies the adjustment criterion for ($X, Y$) in \graph G. If adjustment faithfulness holds, the two are equivalent: $\hz$ is true if and only if $\graph G\vdash \hz$.

We are interested in identifying the most likely adjustment set for \var X, \var Y. Unless otherwise mentioned, when we say that \vars Z is an adjustment set, we mean it is so for \var X, \var Y. Thus, we want to find the set that maximizes the the posterior
\begin{equation}\label{eq:scoreHzDeDo}
P(\hz|D_{exp}, D_{obs})\propto P(D_{exp}|D_{obs},\hz)P(\hz|D_{obs})
% \frac{P(D_{exp},D_{obs}|\hz)P(\hz)}{P(D_{exp},D_{obs})}=\\
% \frac{P(D_{exp}|D_{obs},\hz)P(\hz|D_{obs})}{P(D_{exp}|D_{obs})}
\end{equation}
The score decomposes into (a)  the probability of the experimental data given the observational data and given that \vars Z is an adjustment set (or \hz is true), (b) the probability that \vars Z is an adjustment set given the observational data .

\subsection{Estimating $P(D_{exp}|D_{obs},\hz)$}
$D_{exp}$ includes data $D_x$ for each independent atomic intervention $P_{X=x}(Y)$, and therefore $
P(D_{exp}|D_{obs},\hz)$ decomposes as
$\prod_x P(D_{x}|D_{obs},\hz)$.
For each $x$, we can derive $P(D_{x}|D_{obs},\hz)$ on the basis of the adjustment formula: Under \hz, the adjustment formula connects the post-interventional to the observational distribution. Let $\theta_{Y_x} = \{\theta_{y_x}\}$ be set the parameters representing the probabilities $P(Y\Equal y|do(X\Equal x))$. Then, $P(D_{x}|\theta_{y_x}, D_{obs}, \hz\!)$ $\Equal P(D_{x}|\theta_{y_x})$. Integrating over $\theta_{Y_x}$, we have that
\begin{equation}\label{eq:exp_post}
P(D_{x}|D_{obs}, \hz)=
\int_{\theta_{Y_x}} P(D_{x}|\theta_{Y_x})f(\theta_{Y_x}|D_{obs}, \hz)d\theta_{Y_x}.
\end{equation}

$f(\theta_{Y_x}|D_{obs}, \hz)$ represents the posterior density for $\theta_{Y_x}$ given the observational data, if $\vars Z$ is an adjustment set. We use $\theta_\vars z$ denote the parameter for $P(\vars Z=\vars z)$.
%Let $\vars Z$ have $Κ$ unique configurations, $k\in \{1, \dots, K\}$. For simplicity, we use $\mathbf z=k$ to denote that $\vars z$ takes its $k$-th configuration.  We use $\theta_k$ to denote the parameter for $P(\vars z\Equal k)$. 
%For brevity, in the remainder of this paper we present the theoretical results assuming \var Y is a binary variable, taking values $0$ or $1$, which has a Beta distribution. Extension to a multinomial distribution can be seen in  Algorithm \ref{algo:pde}. 

Let $\theta_{y|x, \vars z}$ be the parameters for  $P(Y\Equal y|X\Equal x, \mathbf Z\Equal z)$. Under \hz,  $\theta_{y_x}\Equal \sum_{\vars z}\theta_{y|x, \vars z}\theta_\vars z$  for all $y\in Y$. Let $N^y_{x}$ be the counts where $Y\Equal y$ in $D_{x}$. We can now recast Eq. \ref{eq:exp_post} to include only observational parameters, as follows:
\begin{equation}\label{eq:exp_post_final}
\begin{split}
P(D_{x}|D_{obs}, \hz) =\int_{\theta_{y|x, \vars z}}\int_{\theta_\vars z}
\prod_{y}[(\sum_{\vars z}\theta_{y|x, \vars z}\theta_\vars z)^{N^y_{x}}\\\prod_{\vars z} f(\theta_{y|x, \vars z}, \theta_\vars z|D_{obs}, \hz)]d\theta_{y|x,\vars z}d\theta_\vars z,
\end{split}
\end{equation}
where we use the notation $\int_{\theta_i}()d\theta_i$ to denote multiple integration $\int_{\theta_1}\mydots\int_{\theta_I}()d\theta_1\mydots d\theta_I$. Eq. \ref{eq:exp_post_final} captures the proximity of the ID $\hat P_Y(x)$ in $D_{x}$ to the ID we can estimate from $D_{obs}$ using \vars Z as an adjustment set for $X, Y$. $f(\theta_{Y|x,\vars z}|D_{obs}, \hz) = f(\theta_{Y|x, \vars z}|D_{obs})$ is the posterior density for the parameters $\theta_{Y|x,\vars z}$ given $D_{obs}$. 
%These are parameters in the observational distribution, and therefore independent of \hz given  $D_{obs}$.

Eq. \ref{eq:exp_post_final} has no closed form solution, but we can approximate using a sampling procedure described in Alg. \ref{algo:pde}: The algorithm takes as input a posterior Bayesian Network (BN) $\graph B_{post}$, learnt from the observational data.  $\graph B_{post}$ consists of a DAG graph \graph B and the posterior distributions for its parameters  $f(\theta_{i|pa_i}|\graph B, D_{obs})$. 
This BN will be used to do  Bayesian inference for the observational parameters. Thus, graph \graph B need not (and cannot, since latent confounders are possible) represent the true causal relationships among $\vars V$, it just needs to accurately represent the observational distribution $\graph P$.
We then sample from this set of posteriors (line 3) to obtain an instantiation $\widetilde\theta_{i|pa_i}$ of the BN, and use Bayesian inference (function \BayesInf, Alg. \ref{algo:pde}, line 4) to estimate the parameters $\theta_{y|x,\vars z}, \theta_\vars z$ that are required for adjustment. We then use these parameters to compute the corresponding experimental parameters $\widetilde \theta_{Y_x}$ (line 5), and score the experimental data (line 6). We repeat the process over $niters$ samples, and take the average over all samples.

Under $\mathcal H_{\not \exists}$,  we can not use the adjustment formula to connect $D_{obs}$ to the ID, thus $f(\theta_{Y_x}|D_{obs})=f(\theta_{Y_x}) $, and thus
\begin{equation}\label{eq:neg_exp}
\begin{split}\small
P(D_{x}|D_{obs},\mathcal H_{\not\exists}) =
\int_{\theta_{Y_x}} P(D_{x}| \theta_{Y_x})f(\theta_{Y_x})d\theta_{Y_x}.
\end{split}
\end{equation}
For multinomial distributions, we can compute Eq.\ref{eq:neg_exp} in closed form using a weak uniform prior (Alg. \ref{algo:pde}, line 9). If $P(D_{x}|D_{obs},\mathcal H_{\not\exists})>P(D_{x}|D_{obs},\mathcal H_{\vars Z})$, then $\vars Z$ does not give an estimate closer to the experimental data  than using a weak uniform prior. Thus, $\mathcal H_{\not\exists}$ complements the space of hypotheses with respect to the adjustment criterion. 
%only: It may be possible that  $P_X(Y)$ is identifiable from the observational distribution through other rules of do-calculus (e.g, the front-door criterion).  However, our method is limited to inferences that are based on the back-door adjustment.
\subsection{Estimating $P(\hz|D_{obs})$}
To estimate Eq. \ref{eq:scoreHzDeDo} we also need to estimate $P(\hz|D_{obs})$, i.e., the probability that $\hz$ is true, based on the observational data (function \EstProbObs  in Alg. \ref{algo:findBestAdj}). One way to proceed is to consider \hz based on the causal graphs that are plausible given $D_{obs}$. This requires an additional assumption, similar to faithfulness for the adjustment criterion. Specifically, we need to assume that the adjustment sets for (\var X, \var Y) are exactly those for which the adjustment criterion holds. We call this assumption \textbf{adjustment faithfulness}:
 \begin{definition}Let \graph G be a causal SMCM and \graph P a distribution faithful to \graph G over a set of variables $\vars V$, and $X, Y\in \vars V$. Then $\vars Z\subset \vars V\setminus \{X, Y\}$ is an adjustment set for (\var X, \var Y) in \graph P  only if \vars Z satisfies the adjustment criterion for (\var X, \var Y) in  \graph G.
 \end{definition}
Let $\graph G\vdash \hz$ denote that $\vars Z$ satisfies the adjustment criterion for ($X, Y$) in \graph G. If adjustment faithfulness holds, $\hz$ is true if and only if $\graph G\vdash \hz$. Under adjustment faithfulness, we can consider $P(\hz|D_{obs})$  in the space of possible SMCMs:
\begin{equation}\label{eq:probDogivHwBCCD}
\begin{aligned}
P(\hz|D_{obs})=
\frac{\sum_{\graph G\vdash \hz} P(D_{obs}|\graph G)P(\graph G)}{\sum_{\graph G}P(D_{obs}|\graph G)P(\graph G)}
\end{aligned}
\end{equation}
Eq. \ref{eq:probDogivHwBCCD} requires exhaustive enumeration of all possible graphs, and a method for obtaining the posterior probability of an SMCM given the data, both of which are complicated. For large sample sizes,  the true Markov equivalence class $[\graph G]$ will dominate this score. Assuming our sample size is large enough that we can obtain $[\graph G]$ using a sound and complete algorithm like FCI, we can use Eq.  \ref{eq:probDogivHwBCCD} with $P(\graph G) =1$ if $\graph G \in [\graph G]$, and  $P(\graph G) =0$ otherwise. This still requires  enumeration of all the possible members of  [\graph G], which can be done with a logic-based method for learning causal structure  \citep[e.g., ][]{triantafillou2015}. We have developed a method that encodes the invariant features of $[\graph{G}]$ and the adjustment criterion in Answer Set Programming \citep[ASP]{asp2011}. We can then query the logic program for all sets where \hz (does not) hold(s), and use the number of models to compute Eq. \ref{eq:probDogivHwBCCD}. We call the method Graphical Approach (GA). Details and proof of its soundness can be found in the Supplement.

GA has very limited scalability. A more graph-agnostic method is to consider variables that are correlated with both $X$ and $Y$ as possible members of an adjustment set. Specifically, let $\vars Z_{XY}$ be the set of variables that are statistically dependent with both \var X and \var Y. Then, we consider all subsets of $\vars Z_{XY}$ equally probable adjustment sets given $D_{obs}$. 
In experiments in random networks with 5 observed and 5 latent variables, we found that the choice of these two methods for computing \EstProbObs  does not affect the behavior of the algorithms. This result is expected, since the impact of $P(\hz|D_{obs})$ shrinks with increasing experimental samples. We therefore use the more efficient, non-graphical approach in the rest of this work.

\subsection{Finding Optimal Adjustment Sets\label{sec:selection}}
To select the most probable adjustment set, we use  Alg. \ref{algo:pde} to score different adjustment sets $\vars Z$,  and select  $\vars Z^\star\Equal  argmax_{\vars z}{P(D_{exp}|D_{obs},\hz)P(\hz|D_{obs})}$. Notice that the adjustment hypotheses are not necessarily mutually exclusive; multiple sets can be adjustment sets for $(X, Y)$, and explain the observational data equally well; thus, we may have multiple optimal solutions $\mathbf Z^\star$, but they all lead to the same ID.

Alg. \ref{algo:findBestAdj} (FAS) describes the process of selecting an optimal adjustment set: The algorithm takes as input a set of observational data $D_{obs}$ over variables \vars V and a collection of experimental data $D_{exp} =\{D_x\}$ that measure the $Y$ under different manipulations $do(X\Equal x)$. The algorithm initially  learns a posterior BN from the observational data, and forms the set of possible adjustment variables $\vars Z_{XY}$,  by keeping all variables associated with both \var X and \var Y. This set is a superset of at least one true adjustment set, if one exists (Proof in the Supplement), so FAS will asymptotically score  at least one true adjustment set. Subsequently, the algorithm obtains $P(D_{exp}|D_{obs}, \hz)P(\hz|D_{obs})$ for all subsets of $\vars Z_{XY}$, as well as $\mathcal H_{\nexists}$, and returns the best-scoring set (or $\nexists$).

The method also returns an estimate $P_X(Y)$ based on the optimal adjustment set, computed as the average estimate over all sampling iterations. If $\nexists$ is selected, the method returns the experimental estimate, as it has found no adjustment set that can improve it. 

In the worst case, the complexity of the algorithm is exponential in the number of variables, since \LearnBN and \BayesInf are NP-hard problems, and the number of possible subsets increases exponentially with the number of variables. However, we can restrict \LearnBN and \BayesInf only to the variables in $\vars Z_{XY}$. The main factor in the scalability of the method is the number of variables need to consider for adjustment. 
\subsection{Dealing with selection in the experimental data.}
So far, we have assumed that the observational and experimental data are sampled from the same population. This is true in some settings, like biological experiments and point-of-care trials. In most clinical settings,
however, the populations may differ due to inclusion/exclusion criteria, or background differences in the populations (e.g., age distributions due to geographical location). The inclusion/exclusion criteria 
are always reported in an RCT study. In addition, the marginal distributions of some covariates are reported (usually in ``Table 2" of the publication). 

When the RCT trial has been performed on a different population, the corresponding ID cannot be computed using adjustment from $D_{obs}$, since both $P(y|x, z)$ and $P(z)$ may be different in this population. This also means that the effect in the RCT may not be valid for our observational population. In this section, we generalize our method to situations where the RCT is performed on a different population, as described above, utilizing the information in an RCT publication. The method models the differences in the RCT population as selection in the RCT population, and constructs a BN that captures this selected observational distribution.  FAS can then be applied using this selection BN. If it identifies an adjustment set, we can use it to get an unbiased ID estimate for the observational population.

\begin{figure}[t]\label{example1}
\centering
\begin{tabular}{cc}
\includegraphics[width =.4\columnwidth]{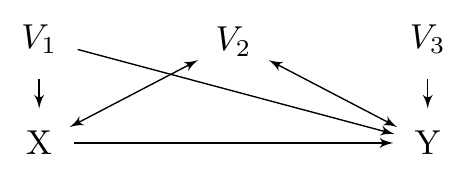}&
\includegraphics[width =.4\columnwidth]{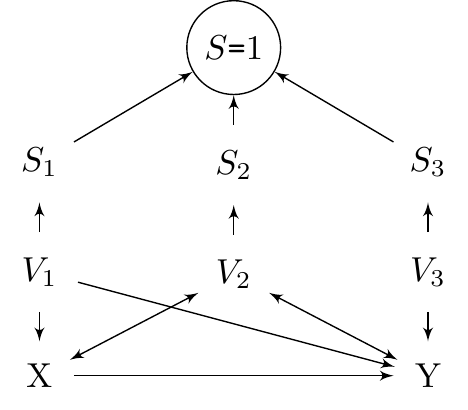}\\
$\graph G$ & $\graph G_{S\Equal 1}$
\end{tabular}

\caption{\label{fig:selection}\textbf{Modeling selection in the experimental population.} \graph G is the true graph for the observational population, and $\graph G_{S\Equal 1}$ corresponds to a selection mechanism in $D_{exp}$ (before randomization), based on variables $\vars V_S={V_1, V_2, V_3}$.  Each variable in $\vars V_{S}$ is selected through a mechanism $P(S_i=1|V_i)$. The mechanisms are mutually independent. $S$ is a binary variable that denotes inclusion in the experimental population, and is true when all $S_i$ are true. This model can describe individual selection criteria, as well as differences in background distributions of individual covariates.} 
\end{figure}
We assume that there is no selection bias in our observational data $D_{obs}$. We also assume that the randomization is performed on a selected population: Specifically, we assume a subset of pre-treatment variables $\vars V_{S}\subset \vars V\setminus \{X, Y\}$ have been selected upon. In this work, we assume that all the selected variables are included in the experimental study: $\vars V_{S}\subseteq\vars  V_{exp}\subseteq \vars V\setminus \{X, Y\}$, and that the marginal distribution of each selected variable is included in the RCT publication. This is always true for inclusion/exclusion criteria, and often for other covariates, such as demographic variables. Finally, we assume that each variable in $\vars V_S$ is independently selected through some mechanism  $P(S_i=1|V_i)$. For example, if $V_i=v$ is an exclusion criterion, $P(S_i=1|v)=0$. Inclusion in the experimental population is then denoted with a binary variable $S =\bigwedge_i S_i$. Let $P^\star$ be the distribution of the experimental population before randomization,, i.e, $P^\star(\vars V)= P(\vars V|S\Equal 1)$. Figure \ref{fig:selection} shows an example of the assumed selection process, described by $\graph G_{S\Equal 1}$. Notice that the selection process may open some backdoor paths between \var X and \var Y. Therefore, an adjustment set in $\graph G$ is not necessarily an adjustment set in $\graph G_{S\Equal 1}$. However,if a set \vars Z is an adjustment set in $\graph G_{S\Equal 1}$, then \vars Z is an adjustment set in $\graph G$. (See Supplement). If \vars Z is an adjustment set in $\graph G_{S\Equal 1}$, the ID in $D_{exp}$ is
\begin{equation}\label{eq:adj_set_sel}
P^\star_x(Y) = 
\sum_z P(Y|x, \vars z,S\Equal 1)P(\vars z| S\Equal 1)
\end{equation}

Alg. \ref{algo:selection} describes a strategy for estimating $P(Y|x, \vars z,S\Equal 1)$, $P(\vars z|S\Equal 1)$ from $D_{obs}$ and the marginal distributions in $D_{exp}$. The method constructs a BN $\graph B_{S\Equal 1}$ that captures the distribution $P(\vars V|S\Equal 1)$ induced by the true selection SMCM $\graph G_S$. It starts with learning a BN that captures the observational distribution $P(\vars V)$\footnote{This graph can asymptotically be learnt with a  Bayesian marginal likelihood score but not with a constraint-based method \cite{bouckaert1995}.} and then adds the selection variables and estimates parameters for these variables.  For every variable $V_i$ in $D_{exp}$, we add a new binary variable $S_i$ and an edge $V_i\rightarrow S_i$. Finally, we add a new variable $S$, with an edge $S_i\rightarrow S$ for each $S_i$. We call this DAG the \emph{selection DAG}. The parameters are constrained to preserve the marginal distributions in $D_{exp}$ (line 5). The resulting constraint satisfaction problem can be solved with any numerical method. It has infinite solutions, but they all lead to the same distribution $P(\vars V|S\Equal 1)$. The output of the method is a selection BN $\langle \graph B_S, \theta_{\graph B_S}\rangle$ that can capture the pre-intervention distribution $P(\vars V|S\Equal 1)$ of the experimental population. The process is asymptotically correct, in the sense that if the true selection SMCM is $\graph G_{S\Equal 1}$ as described above, $\langle \graph B_S, \theta_{\graph B_S}\rangle$ can be used to estimate $P(\vars V|S=1)$ (Proof in the supplementary). We  can estimate the quantities in Eq. \ref{eq:adj_set_sel} using inference on $B_S$. Notice that we can estimate these quantities for any set $\vars Z$ in $D_{obs}$, even if it includes variables that are not in $\vars V_{exp}$.

\begin{algorithm}[t]
\LinesNumbered
\SetKwInOut{Input}{input}
\SetKwInOut{Output}{output}
\Input{$D_{obs}, D_{exp}$}
\Output{Selection BN $\langle \graph B_S, \theta_{\graph B_S}\rangle$}
$\langle\graph B_S,\hat\theta_{\graph B_S}\rangle \leftarrow \LearnBN(D_{obs})$\;
$C\leftarrow \emptyset$; $\quad/ / $ initialize list of constraints\\
\ForEach{$\var V_i\in \vars V_{exp}$}
{Add $V_i\rightarrow S_i\rightarrow S$ to $\graph B_S$\;
Add the marginal-preserving constraints to $C$: $\sum_{\vars V_{exp}\setminus V_i} \frac{P(\vars V_{exp})\prod_j \theta_{S_j\Equal 1|V_j}}{P(S=1)} = P^\star(V_i)$\;}
Find $\hat \theta_{S_i|V_i}$ that satisfy $C$\;
$\hat \theta_{S=1|\cup_i (S_i=1)}=1$,  $\theta_{S=1|\cup_i S_i}=0$ otherwise\;
$\theta_{\graph B_s}\leftarrow \{\hat \theta_{\graph B_s},\hat \theta_{S_i|V_i}, \hat \theta_{S|\cup_i S_i}\}$
\caption{SelectionBN\label{algo:selection}}
\end{algorithm}

The selection BN can be used in Alg. \ref{algo:pde} instead of $\graph B$ with minimal modifications. Detailed pseudocode for the modified Alg. \ref{algo:pde} for selection bias can be found in the Supplementary. One important difference is that for $\nexists$, the returned estimate for $P_x(Y)$ is N/A instead of the empirical estimate $\hat P_Y(x)$, since this estimate is only valid for the experimental population. Moreover, the proposed method only identifies adjustments sets that are also valid in $\graph G_{S\Equal 1}$. Thus, in this case, $\mathcal H_\nexists$ should be interpreted as ``no adjustment set exists among measured variables that can be used to estimate the ID in the RCT."
When our assumptions are violated and the selected variables in  $\graph G_{S\Equal 1}$ are not included in $\vars V_{exp}$,  (for example, consider $\graph G_s$ if $V_3$ is not reported in $D_{exp}$) we cannot estimate $P(\vars V|S\Equal 1)$ using Alg. \ref{algo:selection}. We expect that our method will  then fail to identify a high-scoring adjustment set (other than by chance) and will return $\mathcal H_\nexists$. In our experiments, under violations of this assumption, the behavior of the algorithm is consistent with our expectation.

\section{Related Work} Identifying causal effects is an important problem and rich literature exists. 
One line of work tries to select an adjustment set from observational data. For the most part, these methods try to select an adjustment set. \citet[henceforth VWS]{vanderweele2011new} propose to control on a set of covariates that satisfy the ``disjunctive set criterion", i.e., adjusting for causes of both the treatment $X$ and the outcome $Y$. The method is guaranteed to adjust for a valid adjustment set, if one exists. However, it requires that we know which variables cause $\var X$ and $\var Y$, while we make no such assumption. \citet[henceforth HPM]{henckel2019graphical} provide methods for selecting an optimal adjustment set for linear Gaussian data with no hidden confounders, when we know all valid adjustment sets. They also provide a pruning method that takes as input a valid adjustment set, and returns a smaller valid adjustment set with lower asymptotic variance\footnote{A similar, but less general,  pruning criterion is presented in \citet{vanderweele2011new}.}. \citet{rotnitzky2020} show that the results hold for broader types of distributions. \citet{smucler2020efficient} extend these some of these results to DAGs with latent variables (though they show that a globally optimal adjustment set may not exist). These methods require that the ground truth graph is known, or that the effect is uniquely identifiable from observational data through covariate adjustment. Thus, in contrast to our FAS, this line of works assumes there is no uncertainty on whether a set \vars Z is an adjustment set. 

Another line of work focuses on identifiability of causal effects based on ME classes of SMCMs: \citet{perkovic2017complete} present algorithms for identifying adjustment sets in a PAG [\graph G], when $P_X(Y)$ is uniquely identified through adjustment in all the graphs of the corresponding ME class. The method otherwise returns that [\graph G] is not ``amenable" relative to the desired effect.
\citet[algorithm LV-IDA]{malinsky2017estimating} compute bounds on causal effects for linear Gaussian models by estimating all the IDs identifiable through adjustment by at least a graph in the ME class of graphs. These sets will include N/A if the effect is not identifiable in at least one graph in the ME class. \citet{jaber2019causal}  and \citet[henceforth HEJ]{hyttinen2015calculus} present complete algorithms for identifying causal effects in a PAG [\graph G] using do-calculus. These methods can also identify some effects not identifiable through adjustment (e.g., the front-door criterion). If an effect is not uniquely identifiable in [\graph G], HEJ can output a list of all possible causal effects (including N/A if the effect is non-identifiable in some $\graph G\in [\graph G]$). All these approaches are  complete for their respective goals for PAGs derived from observational data. In our case, where the causal effect $P_X(Y)$ is also available and assumed to be non-zero, this restricts the ME class [\graph G] to graphs that satisfy all the conditional (in) dependence constraints in $D_{obs}$ and $D_{exp}$ (there is one constraint in $D_{exp}$: the pairwise dependence of $X$ and $Y$). We do not know if the methods are complete in this setting. 

The main difference between FAS and these methods is that they will output a single estimate for $P_X(Y)$ only if all the graphs that are consistent with the constraints in $D_{obs}$ and $D_{exp}$ imply the same estimate. For example, graphs in Fig. \ref{fig:example1} are consistent with the CIs (m-connections/separations) in $D_{exp}$ and $D_{obs}$, but imply  different estimates for $P_D(AE)$ from $D_{obs}$. These methods would return N/A, with the exception of HEJ and LV-IDA thad would return all possible quantities: $P_D(AE)\in \{P(AE|D), \sum_c P(AE|D, c)P(c), N/A\}$. In contrast, our method generates a higher score for the estimate that is closer to the sample estimate $\hat P_D(AE)$, and uses this score to select the most likely adjustment set out of the three.

%\citet{entner13a} propose a rule for identifying adjustment sets, based on the (conditional) independence and dependence relationships among the observed variables. The method is guaranteed to find an adjustment set, if this set holds in all Markov Equivalent graphs consistent with $D_{obs}$. Adjustment sets are not mutually exclusive; multiple sets are often valid for the same effects. 

Some methods also combine experimental and observational data sets in different settings than ours. For continuous data and linear relationships, $D_{obs}$ and limited $D_{exp}$ data can be combined to learn linear cyclic models \citep{eberhardt2010combining}. \citet{kallus2018removing} propose a method improving conditional interventional estimates, but the method requires some overlap of covariates between $D_{obs}$ and $D_{exp}$ data, a binary treatment and continuous covariates and outcome. \citet{rosenman2018propensity} propose combining RCT and observational data to improve causal effect estimates, based on some similar assumptions to in the current paper. However, the method requires the event-level RCT data, and assume no hidden confounders.  \citet{wangcost} combine observational and limited experimental data, but focus on identifiability of causal effects and assume no hidden variables.

For the task of generalizing causal effects across different populations with selection bias, \citet{bareinboim2013general} and \citet{correa2018generalized} present general identifiablity results when the true graph is known.

\section{\label{sec:experiments}Experiments}
\begin{figure*}[th!]
\centering
\begin{tabular}{cccc}
&(a) \textbf{no selection} & (b) \textbf{observed selection} &
(c) \textbf{latent selection}\\% & (d) AUCs\\
\rotatebox[origin=l]{90}{\phantom{ffff}random SMCMs}&\includegraphics[width = 0.25\textwidth]{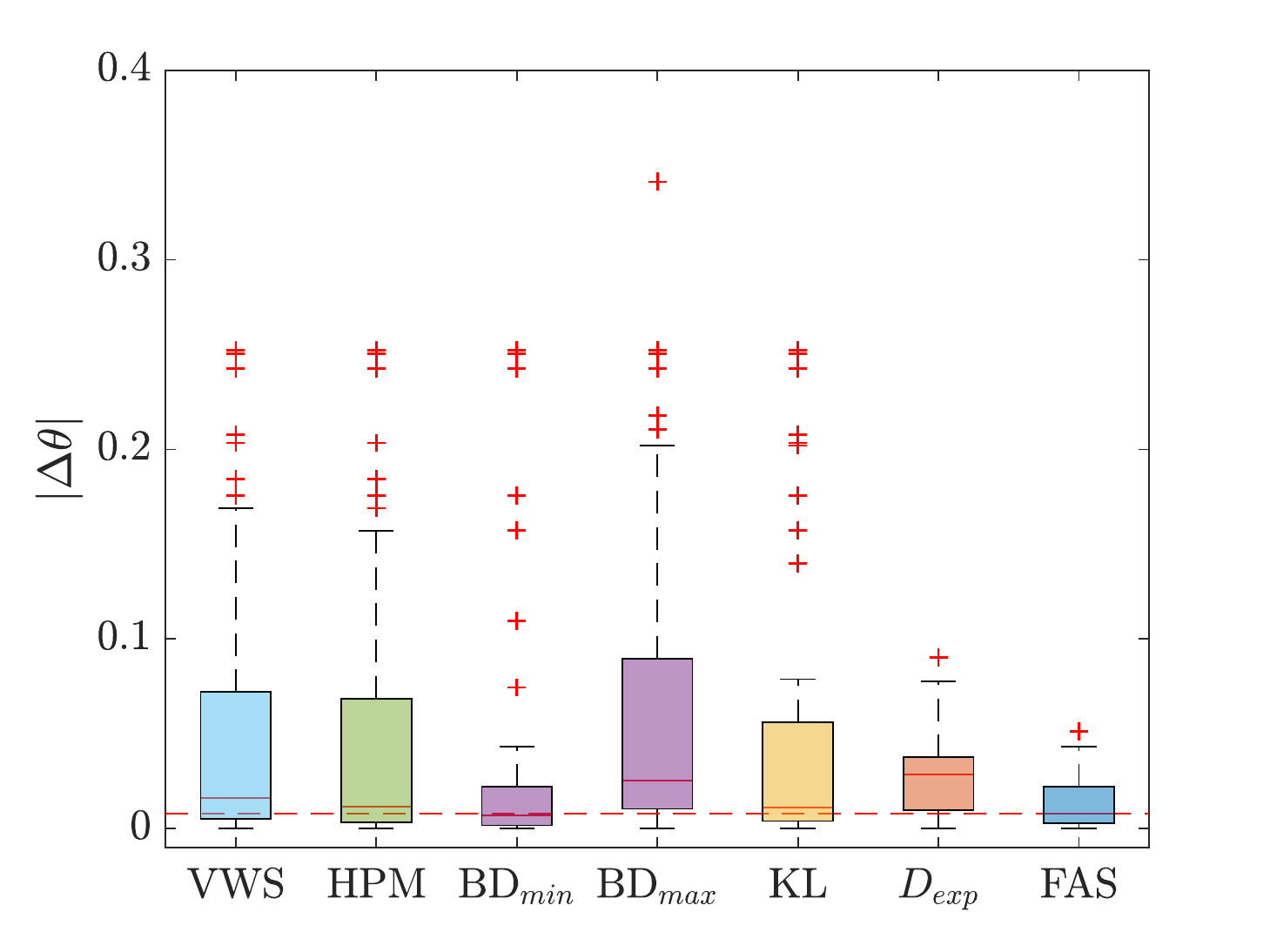}&
\includegraphics[width = 0.25\textwidth]{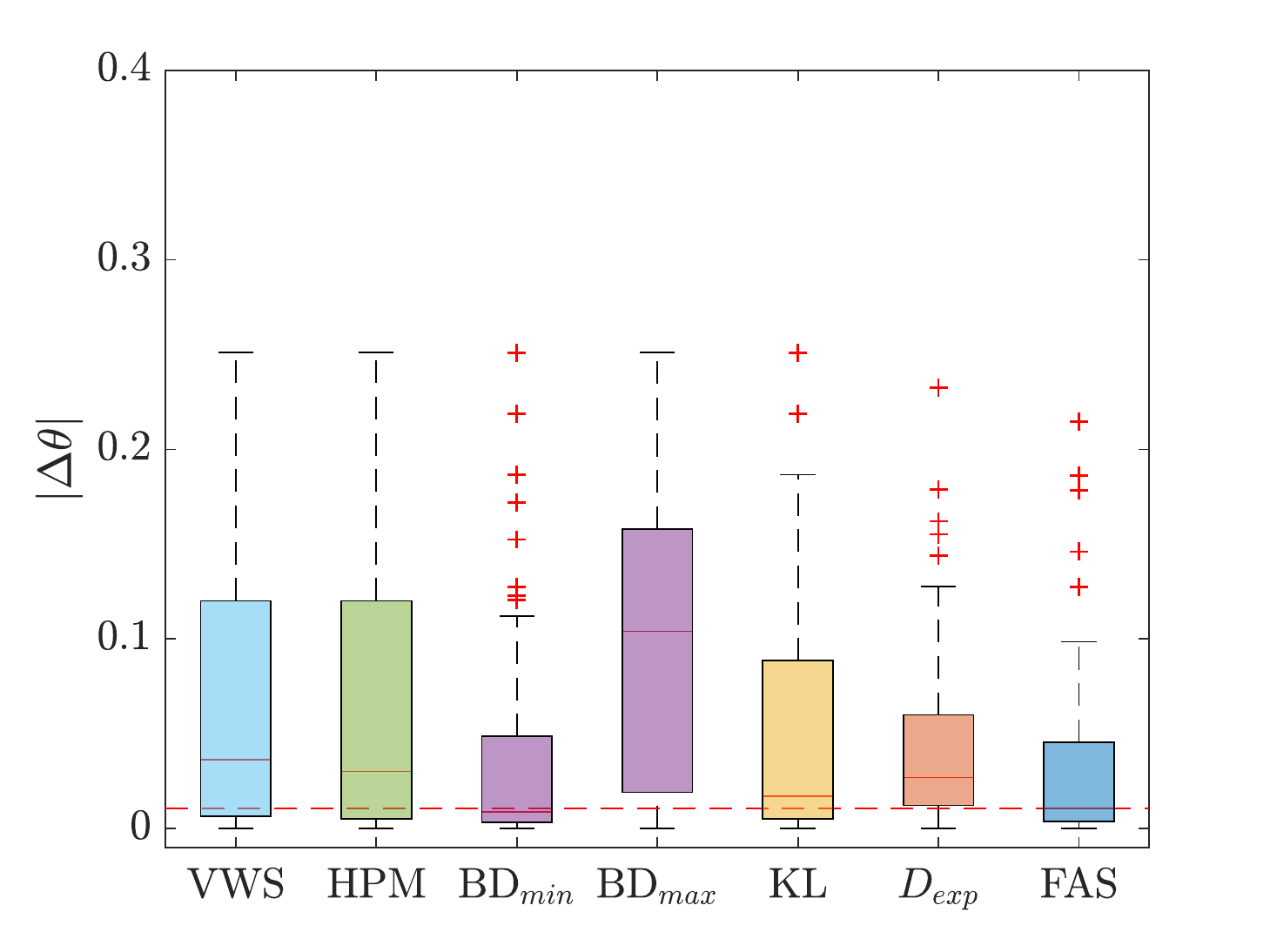}&
\includegraphics[width = 0.25\textwidth]{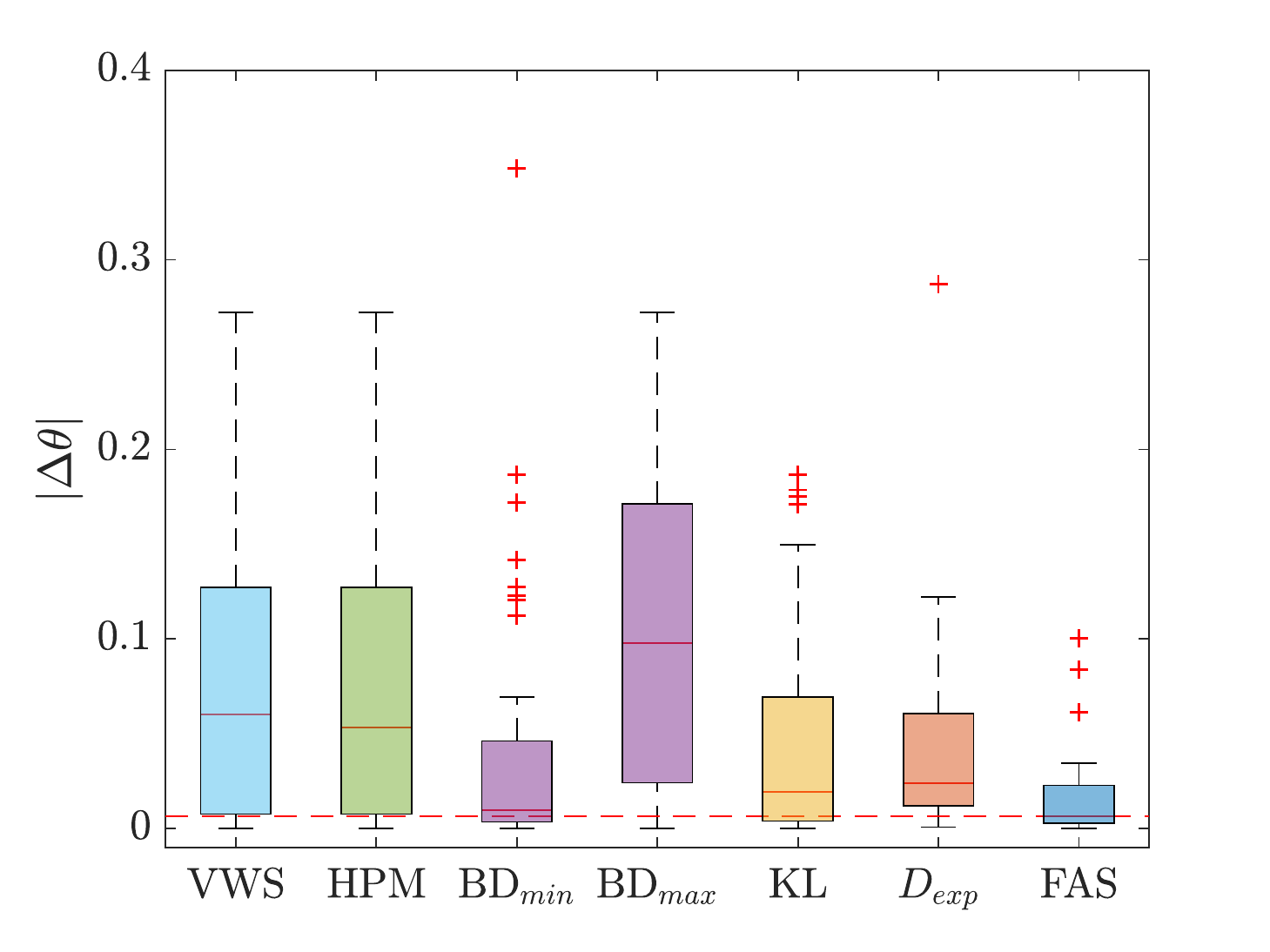}\\
\rotatebox[origin=l]{90}{\phantom{ffff}pretreatment}&
\includegraphics[width = 0.25\textwidth]{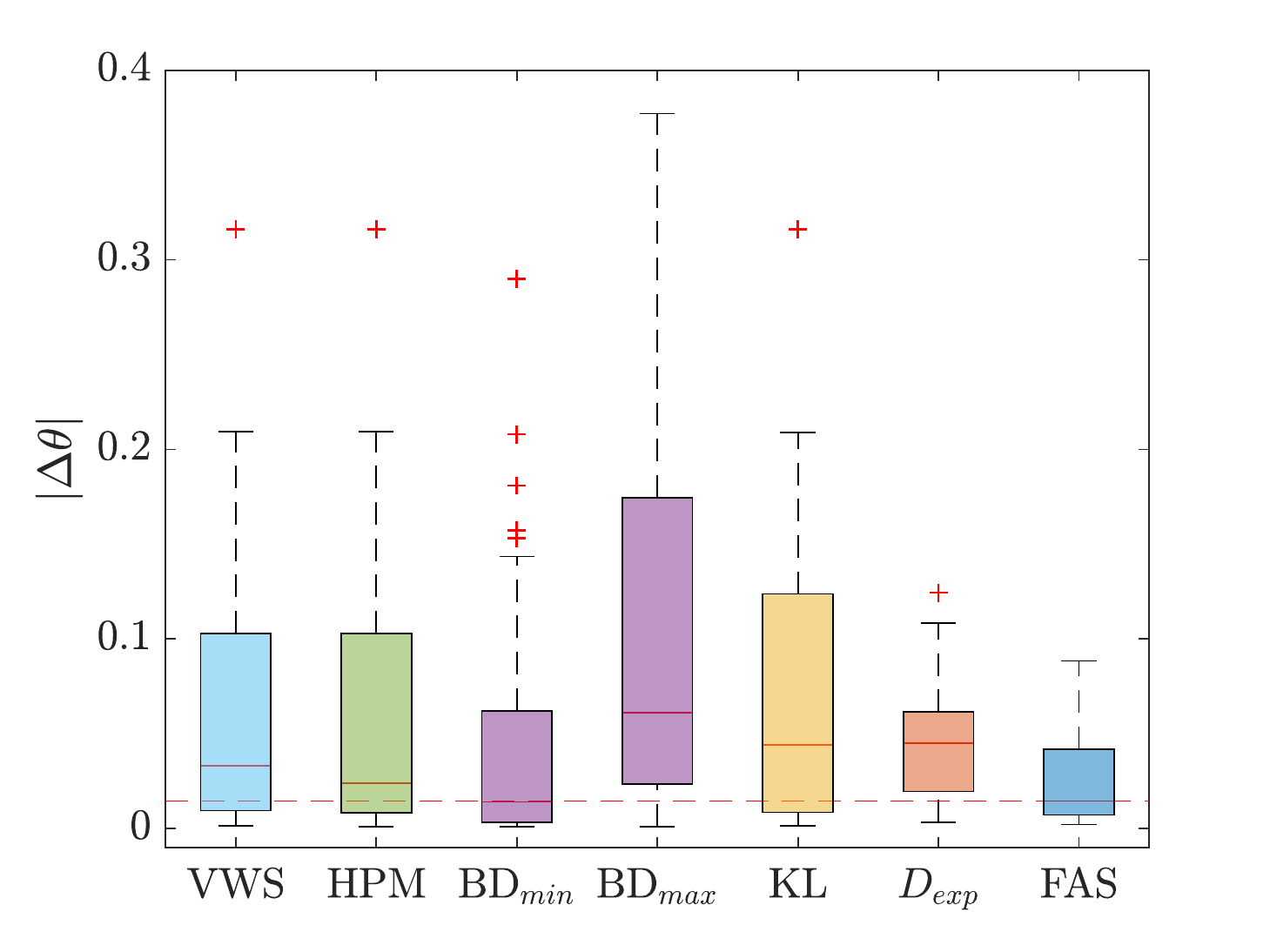}&
\includegraphics[width = 0.25\textwidth]{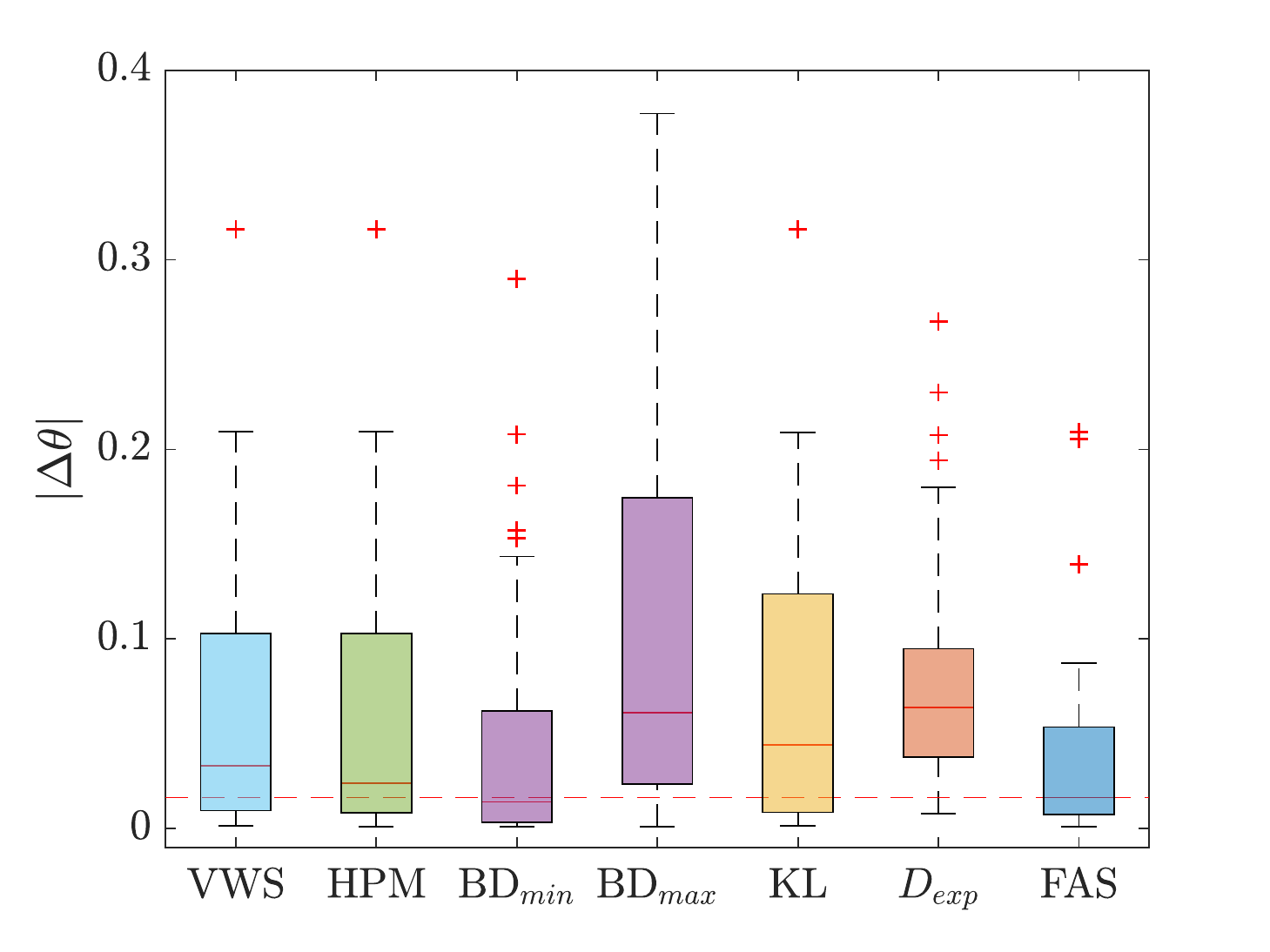}&
\includegraphics[width = 0.25\textwidth]{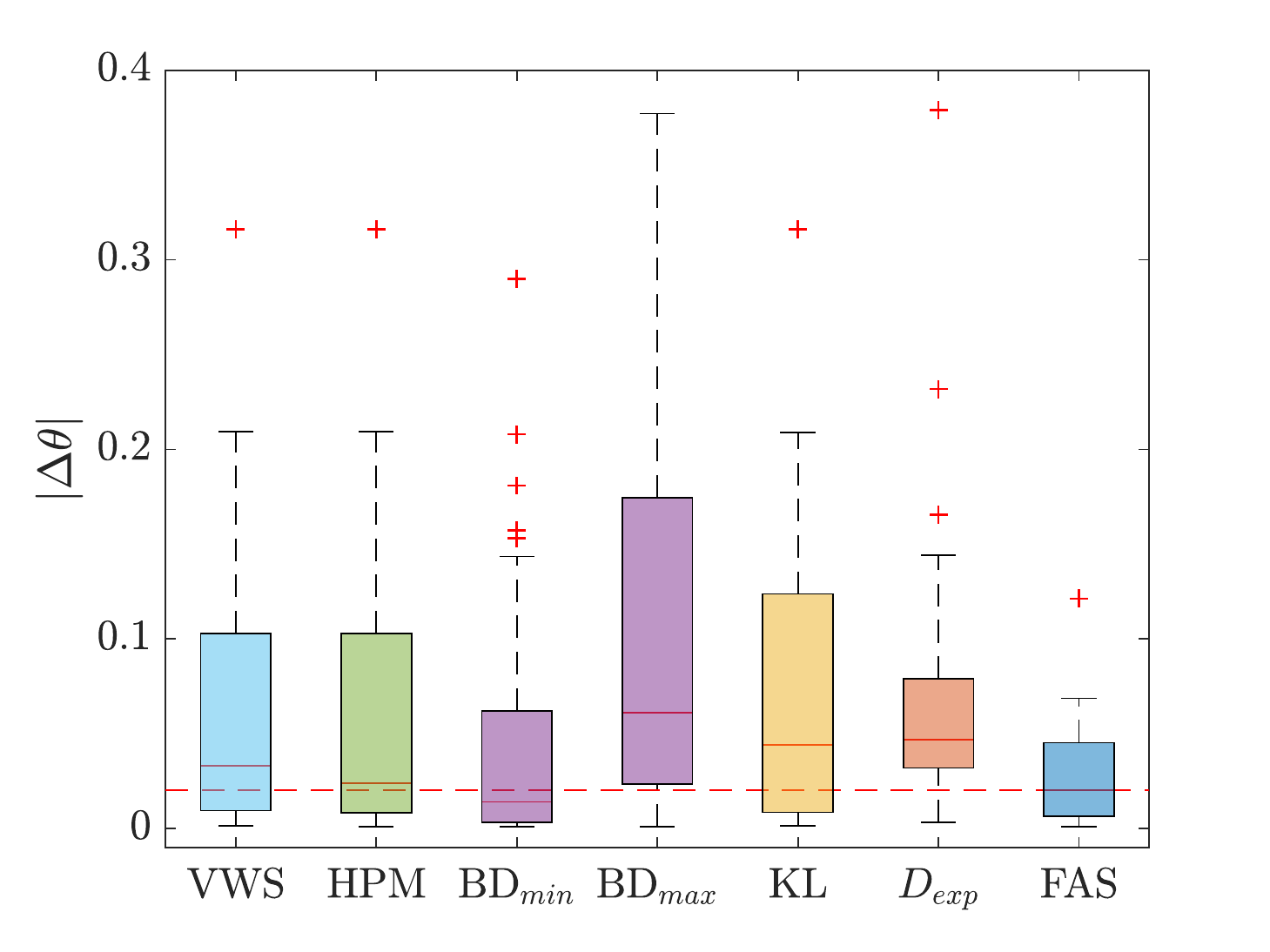}\\
\end{tabular}
\caption{\label{fig:estEffects} Boxplots for the distribution of $|\Delta\theta|$ over 50 iterations in random SMCMs (top row) and when all variables are pre-treatment (bottom row). FAS is shown in blue. Red dotted line corresponds to  FAS median.} 
\end{figure*}

% \begin{figure}[th!]
% \begin{tabular}{cc}
% \includegraphics{figures/pag1.pdf}     &  
% \includegraphics[width =.45\columnwidth]{figures/pag.pdf}\\
% (a)  $\graph G_1$ & (b)  $\graph G_2$
% \end{tabular}
% \caption{\label{fig:cigraph}\textbf{SMCMs with bidirected edges used to simulate data.} In $\graph G_1$ there is no observed adjustment set for \var X, \var Y. In $\graph G_2$, \var Z is the only adjustment set for $X, Y$. In both cases, the (absence of an) adjustment set is not identifiable in the ME class defined by $D_{obs}$ and $D_{exp}$.} 
% \end{figure}

% \begin{figure*}[th!]
% \begin{tabular}{ccc}
% \includegraphics[width =.3\textwidth]{figures/smcm.pdf}&
% \includegraphics[width =.3\textwidth]{figures/smcm_dist.pdf}\\
%  & 
% (b)Ratio of selection&
% (c)&
% \end{tabular}
% \caption{\label{fig:cigraphres}(a). Dashed edges are means, solid edges are medians. (b-c)\textbf{Alg. \ref{algo:findBestAdj} can correctly identify causal effects} (b)Fraction of times where Alg. \ref{algo:findBestAdj} returns each possible adjustment set for data simulated from the graph in Fig. \ref{fig:example1}c. Alg. \ref{algo:findBestAdj} identifies that we must control for \var Z, but sometimes includes \var W in $\vars Z^*$, when the bias from adjusting for a collider is small. (c)  Distribution differences. In all cases, our method improves the estimation of $P(Y|do(X))$.} 
% \end{figure*}
\textbf{Simulations setup.} We examined the performance of our method in three different settings: (a) \textbf{no selection:} $D_{obs}$ and $D_{exp}$  are sampled from the same population, (b) \textbf{observed selection:}  $D_{exp}$ is sampled from a selected population, and the marginal distribution $P(V|S=1)$ of each selected variables are included in $D_{exp}$, and (c) \textbf{latent selection:}  $D_{exp}$ is sampled from a selected population, but the selected variables are not reported in $D_{exp}$. We simulated $D_{obs}$ with 10,000 samples from  DAGs with mean in-degree 2. Each DAG includes a pair $\var X\dashrightarrow \var Y$, and 10 additional covariates: 6 observed and 4 latent. We used two types of DAGs: (i) random DAGs  and (ii) DAGs where all the additional covariates are pre-treatment. Variables were discrete  with 2-3 categories each and random parameter values $P(\vars X|Pa(X))$. A random subset of the observed variables were  included in the experimental data (their marginal distributions are reported in the experimental data).  Selection bias was imposed by adding binary selection nodes $S_i$ and random parameters $P(S_i\Equal 1|V_i)$. For \LearnBN, we used FGES \cite{chickering2002optimal} with the default parameters. We used $niters\Equal100$. \textbf{Evaluation measures.} We examined the performance of our algorithms in terms of  their ability to improve causal effect estimation for the observational population: We estimated the absolute distance of the predicted vs the true interventional distribution for the observational population, $|\Delta\theta|\Equal |\hat \theta_{Y_x}-\theta_{Y_x}|$ averaged over all parameters $\theta_{Y_x}$. \textbf{Comparison to other approaches.} We are unaware of any other method designed for these specific settings. We compare against the following: (1) VWS, light blue in Fig \ref{fig:estEffects}: The ``disjunctive criterion" in VWS. The method requires that we know which variables cause \var X and \var Y. We used the ground truth DAG to obtain that information, and  only kept observed variables. (2) HPM: We used the pruning method in HPM to prune the VWS estimate, as this is shown to remove "overadjustment" variables and improve estimates. (3), minimum and maximum of this range in purple in Fig. \ref{fig:estEffects}: This range corresponds to the set of all possible causal effects obtainable through with covariate adjustment, based on the ground truth ME class $[\graph G]$ of SMCMs consistent with both $D_{obs}$ and $D_{exp}$ (obtained using a CI oracle). If in some \graph G in $[\graph G]$, $\mathcal H_\nexists$  holds, then [BD] includes N/A. Asymptotically, this set is  properly included in the set returned by HEJ, since we only include estimates identifiable through the backdoor criterion. The set is also asymptotically what LV-IDA would return. In our simulations, N/A was included in the output (i.e., the ``no adjustment set hypothesis" could not be rejected based on the ME class) in 92 out of 100 total simulation graphs. We report the minimum and maximum of this range, regardless of whether N/A is included in the output. BD$_{min}$ corresponds to the best possible estimate we could get for $P_X(Y)$ by adjusting for observed covariates in these simulations. (4) KL, in yellow in Fig \ref{fig:estEffects}. Instead of computing a Bayesian score for $P(\hz|D_{exp}, D_{obs})$ we identify the set $\vars Z$ that minimizes the KL divergence of the corresponding predicted ID and the empirical ID. However, notice that KL cannot select $\mathcal H_\nexists$. (5) $D_{exp}$, in orange in Fig \ref{fig:estEffects}: Empirical estimate  $\hat P_X(Y)$  in $D_{exp}$. \textbf{Results: } Fig \ref{fig:estEffects} shows results for random SMCMs (top row), and for SMCMs where the covariates are known to be pre-treatment. FAS improves the estimation of $P_X(Y)$($|\Delta \theta|$ closer to zero,  lower variance) compared to $D_{exp}$, particularly in cases in where the experimental data come from a selected population. Despite the fact that VWS and HPM are constructed based on ground truth knowledge that is typically not available, the methods  perform worse than FAS, since they do not utilize the experimental data, and always select an adjustment set, even if none exists in the ground truth structure. In addition, the pruning process in HPM does not seem to improve VWS estimate, possibly because the number of covariates is already low. FAS also outperforms BD$_{min}$. This is because FAS can identify cases where the $P_X(Y)$ is not identifiable from $D_{obs}$ (e.g., \var X and \var Y share a latent confounder). It therefore heavily biased estimates. In the latent selection setting, FAS returned $\graph H_\nexists$ in $22$ out of $50$  cases in random SMCMs (and in $15$ of $50$ in SMCMs with pretreatment only covariates). Thus, when the effect of latent selection is significant, FAS  often  acts conservatively and does not return an adjustment set. 
Average running time for one iteration of the algorithm was 7.95 $\pm$ 12.8 seconds. In the supplementary, we show results for different $D_{exp}$ and $D_{obs}$ sample sizes, different number of covariates, and  running times. \textbf{Real data:} We applied our method to analyze the causal relationship between statin use and its known adverse effect, myalgia. We used EHR data for  100,000 patients from (hospital name removed for anonymity). We used RCT data from the STOMP trial \cite{stomptrial}, which estimated the effect of statin use on myalgia. The study included 203 treatment and 214 control patients, stratified into age groups. We also included variables representing \emph{age, sex, diabetes, thyroid disorders, and hyperlipidemia}. Diabetes and thyroid disorders were exclusion criteria in the study\footnote{The study had some additional exclusion variables which we did not model because they are extremely rare. }. FAS returned $\mathbf Z^\star = \{\textnormal{Age}\}$ as the most likely adjustment set. If we remove age from the covariates, the method returns $\graph H_\nexists$. It is  clear that age is a confounder in this example. However, our method identified it without any prior clinical knowledge of the causal relationships among the modeled variables.
\section{Discussion}
We present a method for learning adjustment sets and improving the estimation of causal effects by combining large observational and limited experimental data (e.g., combining EHR and RCT data). Our results  show that the method can make additional inferences relative to existing methods. Directions for future work include improving scalability, mixed types of data, and generalizations to broader types of selection settings.
\bibliography{adj_set_bib.bib} 

\end{document}